# Empirical Bayes methods for controlling the false discovery rate with dependent data

**Weihua Tang**[1] **and Cun-Hui Zhang**[1,*]

*Rutgers University*

**Abstract:** False discovery rate (FDR) has been widely used as an error measure in large scale multiple testing problems, but most research in the area has been focused on procedures for controlling the FDR based on independent test statistics or the properties of such procedures for test statistics with certain types of stochastic dependence. Based on an approach proposed in Tang and Zhang (2005), we further develop in this paper empirical Bayes methods for controlling the FDR with dependent data. We implement our methodology in a time series model and report the results of a simulation study to demonstrate the advantages of the empirical Bayes approach.

## 1. Introduction

The false discovery rate (FDR), proposed by Benjamini and Hochberg ([1], BH hereafter), has been widely used as an error measure in multiple testing problems. Let $R$ be the number of rejected hypotheses (discovered items) and $V$ be the number of falsely rejected hypotheses, the FDR is defined as

$$\text{(1.1)} \qquad \text{FDR} \equiv E(V/R) I_{\{R>0\}}.$$

Since most discovered items truly contain signals when the ratio $V/R$ is small, FDR controlling methods allow a scientific inquiry to move ahead from a screening experiment to more focused systematic investigations of discovered items. Moreover, since FDR controlling methods do allow a small number of items without signal to slip through, they often provide sufficiently many discovered items in such screening experiments. Thus, the FDR seems to strike a suitable balance between the needs of multiple error control and sufficient discovery power in many applications (e.g. microarray, imaging, and astrophysics studies), compared with the more conservative family-wise error rate (FWER) $P(V > 0)$ and more liberal per-comparison error rate (PCER) $EV/m$, where $m$ is the total number of hypotheses being tested. Mathematically, we notice that

$$\text{PCER} \leq \text{FDR} \leq \text{FWER}.$$

The FDR is closely related to the positive predictive value (PPV) for diagnostic tests, since

$$\text{(1.2)} \qquad \text{PPV} = 1 - V/R$$

*Research partially supported by National Science Foundation Grants DMS-0405202, DMS-0504387 and DMS-0604571.

[1]Department of Statistics, Hill Center, Busch Campus, Rutgers University, Piscataway, New Jersey 08854, USA, e-mail: weihuat@stat.rutgers.edu; czhang@stat.rutgers.edu

based on ground truth.

Most papers in the FDR literature have focused on procedures for controlling the FDR based on independent test statistics or properties of such procedures with dependent test statistics. For example, in BH [1], the $p$-values for individual hypotheses are assumed to be independent and uniformly distributed under the null. Benjamini and Yekutieli [3] proved that the BH [1] procedure still controls the FDR under a "positive regression dependence" condition on test statistics under the null hypotheses. Storey, Taylor and Sigmund [?] proposed certain modification of the BH [1] procedure and proved its FDR controlling properties under conditions on the convergence of the empirical processes of the $p$-values. For related work in controlling decision errors in multiple testing, see Cohen and Sackrowitz [4, 5], Finner and Roters [9], Sarkar [11] and Simes [12], in addition to references cited elsewhere in this paper.

In Tang and Zhang [15], we showed that the optimal solution of the problem of

(1.3) $\qquad$ maximizing $ER$ subject to $E(V/R)I_{\{R>0\}} \leq \alpha$

may produce undesirable multiple testing procedures, and proposed Bayes and empirical Bayes (EB) approaches under a conditional version of (1.3) given certain test statistics which determine $R$. This means to maximize the total discovery $R$ (i.e. power) of multiple testing procedures subject to a preassigned level of the conditional FDR:

(1.4) $\qquad$ $\text{CFDR}(\mathbf{X}) \equiv E\big[V/(R \vee 1)\big|\mathbf{X}\big],$

for certain statistics $\mathbf{X}$ satisfying $E[R|\mathbf{X}] = R$, where $(x \vee y) = \max(x, y)$. These approaches, which provide a general framework for controlling the FDR with dependent data, are discussed in Section 2.

We note that the concept of conditional FDR (1.4) allows conditioning on a mixture of observations, parameters, and variables generated by statisticians to implement randomized rules. The CFDR (1.4) becomes posterior FDR when $\mathbf{X}$ is the vector of all observations of concern. If we observe $\mathbf{X}_0$ and generate variables $\varepsilon_0$ to execute a randomized multiple testing rule, the constraint $E[R|\mathbf{X}] = R$ demands $\mathbf{X} = (\mathbf{X}_0, \varepsilon_0)$ in our optimization problem, cf. (2.1) below, but the posterior FDR is computed given $\mathbf{X}_0$ alone for such randomized tests.

In Section 3, we develop EB methods for controlling the (conditional) FDR in a time series model based on our approach. In Section 4, we present simulation results which demonstrate the advantages our method compared with the BH [1] rule based on marginal $p$-values and the additional knowledge of the proportion of true hypotheses.

## 2. Bayes and empirical Bayes approaches

Let $H_1, \ldots, H_m$, be the null hypotheses to be tested and

$$\theta_i = I\{H_i \text{ is not true}\}, \quad \delta_i = I\{H_i \text{ is rejected}\}.$$

In a Bayes approach, we treat $\theta = (\theta_1, \ldots, \theta_m)$ as a random vector and assume that for certain observations $\mathbf{X}$ (not necessarily all observations), the joint distribution of $\{\theta, \mathbf{X}\}$ is known given the knowledge of the joint prior distribution of $\theta$ and all nuisance parameters (if any). Consider the problem of

(2.1) $\quad$ maximizing $R$ subject to $E\big[V/(R \vee 1)\big|\mathbf{X}\big] \leq \alpha$ and $E[R|\mathbf{X}] = R,$

where $R = \sum_{i=1}^{m} \delta_i$ and $V = \sum_{i=1}^{m} \delta_i(1-\theta_i)$. Let $\pi_i = P\{\theta_i = 0|\mathbf{X}\}$ be the posterior probability of $H_i$ given $\mathbf{X}$. The Bayes rule, which solves (2.1), is given by

$$(2.2) \quad \delta_{(i)} = I\{i \leq k^*\}, \quad k^* \equiv k^*(\alpha) \equiv \max\left\{k \leq m : \frac{1}{k}\sum_{i=1}^{k} \pi_{(i)} \leq \alpha\right\},$$

with the convention $\max \emptyset \equiv 0$, where $\{(1), \ldots, (m)\}$ is the ordering of $\{1, \ldots, m\}$ determined by $\pi_{(1)} \leq \cdots \leq \pi_{(m)}$. The above Bayes solution is optimal for any type of data as long as the joint distribution of $(\mathbf{X}, \theta)$ is available.

The Bayes rule (2.2) provides the most powerful solution for controlling the (conditional) FDR in the sense of (2.1) with general dependent test statistics for general null hypothesis, provided the knowledge of the joint prior distribution of unknown parameters and the computational feasibility for the conditional probabilities of the hypotheses $H_i$ given $\mathbf{X}$. The Bayes rule yields $R = k^*$, since it rejects $H_i$ iff $\pi_i \leq \pi_{(k^*)}$. It clearly controls the FDR at level $\alpha$, since the constraint in (2.1) is stronger than $\text{FDR} = E[V/(R \vee 1)] \leq \alpha$.

In applications where the full knowledge of the joint prior distribution is not available or the computation of the posterior probability of the null hypotheses given data is not feasible, the Bayes rule (2.2) motivates empirical Bayes rules of the form

$$(2.3) \quad \delta_{[i]} = I\{i \leq \widehat{k}\}, \ \widehat{k} = \max\left\{k : \frac{1}{k}\sum_{i=1}^{k} \widehat{\pi}_{[i]} \leq \alpha\right\},$$

where $\widehat{\pi}_i$ are estimates of the posterior probability $\pi_i = P\{H_i|\mathbf{X}\}$ and $\{[1], \ldots, [m]\}$ is an estimate of the ordering $\{(1), \ldots, (m)\}$ in (2.2). The performance of the Bayes rule serves as a benchmark, as the goal of the EB (2.3) here is to approximately achieve optimality in the sense of (2.1). This EB approach is applicable for dependent data as long as suitable estimates $\widehat{\pi}_i$ can be obtained. The ordering $[i]$ can be derived from $\widehat{\pi}_i$ if the ordering $(i)$ is not a known functional of the data.

Tang and Zhang [15] proposed the above Bayes and empirical Bayes approaches and proved the asymptotic optimality of the BH [1] procedure as EB in the sense of (2.1). In the rest of the section, we discuss implementation of the EB method and some related work on Bayes and EB aspects of FDR problems.

Suppose that the expectation in (2.1) depends on an unknown parameter $\xi$, so that the main constraint in (2.1) becomes $\text{CFDR}(\mathbf{X}, \xi) \leq \alpha$. Let

$$(2.4) \quad \pi_i(\mathbf{X}, \xi) = P\{\theta_i = 0|\mathbf{X}, \xi\}$$

denote the conditional probabilities of the null hypotheses $H_i$ given statistics $\mathbf{X}$ and parameters $\xi$. If $f(\mathbf{X}|\xi)$ belongs to a regular parametric family, we may use an EB rule with an estimate $\hat{\xi}$ (e.g. the MLE) of $\xi$ and $\widehat{\pi}_i = \pi(\mathbf{X}, \hat{\xi})$ in (2.3), or a hierarchical Bayes rule with a prior on $\xi$ and $\pi_i = \int \pi_i(\mathbf{X}, \xi)f(\xi|\mathbf{X})\nu(d\xi)$ in (2.2). If $(\theta_1, X_1), \ldots, (\theta_m, X_m)$ are independent vectors for certain test statistics $X_i$, the conditional probabilities $\pi_i(\mathbf{X}, \xi) = \pi_i(X_i, \xi)$ or the average $k^{-1}\sum_{i=1}^{n} \pi_{(i)}$ of their ordered values in (2.2) may still be estimated sufficiently accurately even for high-dimensional $\xi$, e.g. the asymptotic optimality BH [1] rule as EB. How do we implement (2.3) when (2.4) depends on many components of $\mathbf{X}$ and $\xi$ is high-dimensional? We propose EB rules (2.3) with

$$(2.5) \quad \widehat{\pi}_i = \pi_{m,i}(T_{m,i}(\mathbf{X}), \hat{\xi}), \quad \pi_{m,i}(T_{m,i}(\mathbf{X}), \xi) = P\{\theta_i = 0|T_{m,i}(\mathbf{X}), \xi\},$$

with certain lower-dimensional statistics $T_{m,i}(\mathbf{X})$ which are informative about $\theta_i$. This can be also viewed as the EB version of the approximate Bayes rule

$$(2.6) \quad \delta_{(i)'} = I\{i \leq \widetilde{k}^*\}, \quad \widetilde{k}^* \equiv \max\left\{k \leq m : \frac{1}{k}\sum_{i=1}^{k} \pi_{m,(i)'} \leq \alpha\right\},$$

where $\pi_{m,(i)'}$ are the ordered values of $\pi_{m,i}(T_{m,i}(\mathbf{X}), \xi)$. In applications with time series, image, or networks data, $T_{m,i}(\mathbf{X})$ are typically composed of observations "near" the location of the $i$-th hypotheses $H_i$. The idea is to reduce the dimensionality of the function $\pi_i$ to be estimated. In the time series example in Section 3 and certain models of Markov random fields, $\pi_{m,i}(T_{m,i}(\mathbf{X}), \xi)$ depend on $\xi$ only through a lower-dimensional functional of $\xi$, so that the complexity of the estimation problem is further reduced. The cost of such dimension reduction is the bias

$$b_{m,i} = \pi_{m,i}(T_{m,i}(\mathbf{X}), \xi) - \pi_i(\mathbf{X}, \xi).$$

We note that $Eb_{m,i} = 0$ and $\text{Var}(b_{m,i})$ decreases when we add more variables to the vector $T_{m,i}(\mathbf{X})$.

Efron et al. [6] proposed a different EB approach based on the conditional probability $\text{fdr}(x) = P\{\theta_i = 0 | X_i = x\}$ given a univariate statistic $X_i$, called local fdr, and developed multiple-testing methodologies based on certain estimate of it in mixture models. Efron and Tibshirani [7] and Efron [8] further developed this EB approach using an integrated version of local fdr. These notions of FDR and related quantities have been studied in Storey [13] and Genovese and Wasserman [10]. Sarkar and Zhou (personal communication) recently considered the posterior FDR (i.e. the conditional FDR given all observed data) for a number of multiple testing procedures, including a randomized version of (2.2).

## 3. EB methods in a time series model

In this section, we develop EB methods for controlling the (conditional) FDR in the time series model

$$(3.1) \quad X_i = \mu_i + \epsilon_i,$$

where $\{\epsilon_i\}$ is a stationary Gaussian process (e.g. moving average) with

$$(3.2) \quad E\epsilon_i = 0, \quad \text{Cov}(\epsilon_i, \epsilon_{i+k}) = \gamma(k).$$

Our problem is to test $H_i : \mu_i = 0$ versus $K_i : \mu_i \neq 0$, $i = 1, \ldots, m$, based on observations $\mathbf{X} = (X_1, \ldots, X_m)$. We assume that the null distribution of $X_i \sim N(0, \gamma(0))$ is known. We set $\gamma(0) = 1$ without loss of generality.

We derive an EB procedure (2.3) of the form (2.5) under a nominal mixture model in which $(\theta_i, \mu_i)$ are iid vectors with

$$(3.3) \quad \mu_i \big| (\theta_i = 0) = 0, \quad \mu_i \big| (\theta_i = 1) \sim N(\eta, \tau^2)$$

for certain unknown $(\eta, \tau^2)$. We assume $\sum_k \gamma^2(k) < \infty$, which allows us to take advantage of the diminishing correlation. This leads to

$$(3.4) \quad T_{m,i}(\mathbf{X}) = (X_j, |j - i| \leq k), \quad \xi = (\eta, \tau^2, w_0, \gamma(1), \ldots, \gamma(k)),$$

in (2.5), where $w_0 = P\{\theta_i = 0\}$.

As we have mentioned in Section 2, conditioning on the lower-dimensional statistics $T_{m,i}(\mathbf{X})$ is helpful in two important ways: computational feasibility and reduction of the set of nuisance parameters involved. Since the components of $\mathbf{X}$ are all correlated, the posterior probability $\pi_i(\mathbf{X}, \xi)$ in (2.4) and thus the Bayes rule $k^*$ in (2.2) demand the inversion of $m$-dimensional conditional covariance matrices and summation over $2^m$ possible values of $\theta$. Thus, the Bayes rule is computationally intractable. Exact computation of $\pi_{m,i}(T_{m,i}(\mathbf{X}), \xi)$ in (2.5) is much more manageable since it involves the inversion of $(2k+1)$-dimensional covariance matrices and summation over $2^{2k+1}$ possible values of $T_{m,i}(\theta) = (\theta_j, |j-i| \le k)$. Also, the conditional probability $\pi_{m,i}(T_{m,i}(\mathbf{X}), \xi)$ of $H_i$ given $T_{m,i}(\mathbf{X})$ does not depend on $\gamma(j)$ for $j > k$, so that the dimensionality of $\xi$ is much smaller than the sample size if we choose $k = o(m)$.

Given $w_0$, we estimate $\xi$ as follows:

$$(3.5) \qquad \widehat{\eta} = \frac{1}{m} \sum_{i=1}^{m} X_i/(1-w_0),$$

$$(3.6) \qquad \widehat{\tau}^2 = \frac{1}{m} \sum_{i=1}^{m} \frac{X_i^2 - 1}{1 - w_0} - \sum_{1 \le i \le j \le m, j-i > \rho m} \frac{X_i X_j/(1-w_0)^2}{(1-\rho)^2 m^2/2},$$

$$(3.7) \qquad \widehat{\gamma}(j) = \sum_{i=1}^{m-j} \frac{X_i X_{i+j}}{m-j} - \sum_{1 \le i \le j \le m, j-i > \rho m} \frac{X_i X_j}{(1-\rho)^2 m^2/2},$$

where $0 < \rho < 1$. Estimates (3.5), (3.6) and (3.7) are based on the method of moments, since (3.1), (3.2) and (3.3) imply $EX_i = E\mu_i = (1-w_0)\eta$, $EX_i^2 = 1 + E\mu_i^2 = 1 + (1-w_0)(\tau^2 + \eta^2)$, $EX_i X_{i+j} = \gamma(j) + (E\mu_1)^2$, and $EX_i X_j \approx (E\mu_1)^2$ for large $|j - i|$. In order to reduce the bias [composed of terms involving $\gamma(j-i)$], we use the average of $X_i X_j$ with $|i - j| > \rho m$ in (3.6) and (3.7) to estimate $(EX_i)^2 = (E\mu_i)^2$, instead of $(\sum_{i=1}^{m} X_i/m)^2$. In the simulation study discussed in Section 4, we take $\rho = 0.1$.

For the estimation of the proportion $w_0$ of null hypotheses, we use a Fourier method (Tang and Zhang [15] and Zhang [16]) and its parametric bootstrap version. The Fourier method estimates $w_0$ by

$$(3.8) \qquad \widehat{w}_0^{(F)} \equiv \frac{1}{m} \sum_{i=1}^{m} \psi(X_i; h_m), \ \psi(z; h) \equiv \int h\psi_0(ht) e^{t^2/2} \cos(zt) dt,$$

where $\psi_0$ is a density function with support $[-1, 1]$ and $h_m = \{\kappa(\log m)\}^{-1/2}$ is the bandwidth, $\kappa \le 1$. This estimator is derived from

$$E\left[\psi(X_i; h)\big|\mu_i\right] = \int h\psi_0(ht) e^{t^2/2} \cos(\mu_i t) E e^{it\epsilon_i} dt$$

$$= \int \psi_0(t) \cos((\mu_i/h)t) dt \begin{cases} = 1, \ \mu_i = 0, h > 0, \\ \to 0, \ \mu_i \ne 0, h \to 0+ \end{cases}$$

by Riemann-Lebesgue. For the bootstrap version of (3.8), we generate bootstrap samples of $\mathbf{X}$ under the parameter value of $\widehat{\xi}$ in (3.5)-(3.8) and then estimate $w_0$ by

$$(3.9) \qquad \widehat{w}_0^{(B)} = 2\widehat{w}_0^{(F)} - \overline{\widehat{w}_0^*},$$

where $\overline{\widehat{w}_0^*}$ is the average of the estimator (3.8) based on the bootstrap samples. In the simulation study, we use uniform $[-1,1]$ as $\psi_0$ and $\kappa = 1/2$ for (3.8), and we bootstrap 100 times for (3.9).

The Empirical Bayes procedure rejects the null hypotheses associated with the first $\widehat{k}$ smallest estimated conditional probabilities $\pi_{m,i}(T_{m,i}(\mathbf{X}), \hat{\xi})$, with

$$(3.10) \qquad \widehat{k} = \max\left\{ k : \frac{1}{k} \sum_{i=1}^{k} \widehat{\pi}_{[i]} \leq \alpha \right\},$$

where $\widehat{\pi}_{[1]} \leq \cdots \leq \widehat{\pi}_{[m]}$ are ordered values of $\pi_{m,i}(T_{m,i}(\mathbf{X}), \hat{\xi})$, and $\hat{\xi}$ is defined through (3.5)-(3.8) with the alternative of using the bootstrap estimate (3.9) for $w_0$. The conditional probability

$$\pi_{m,i}(T_{m,i}(\mathbf{X}), \xi) = P\{\theta_i = 0 | T_{m,i}(\mathbf{X}), \xi\},$$

with $\xi$ replaced by $\hat{\xi}$, is computed by conditioning on $T_{m,i}(\theta) = (\theta_j, |j - i| \leq k)$. To save notation, we may drop the subscript $m$ in the rest of the paragraph, e.g. $T_{m,i} = T_i$. Under the nominal mixture model (3.3), the conditional joint distribution of $T_i(\mathbf{X})$ is multivariate normal

$$T_i(\mathbf{X}) \big| T_i(\theta) \sim N\Big(\eta T_i(\theta), \Sigma_i(\theta)\Big),$$

where the covariance matrix $\Sigma_i(\theta) = \mathrm{Cov}\big(T_i(\varepsilon)\big) + \tau^2 \mathrm{diag}\big(T_i(\theta)\big)$ depends on unknown parameters $T_i(\theta)$, $\{\gamma(j) : 1 < j \leq k\}$ and $\tau^2$. The joint density of this conditional distribution is

$$f_i\Big(\mathbf{v}_i \big| T_i(\theta)\Big) = \frac{\exp\big[-(\mathbf{v}_i - \eta T_i(\theta))\{\Sigma_i(\theta)\}^{-1}(\mathbf{v}_i - \eta T_i(\theta))'/2\big]}{(2\pi)^{d_i/2}\{\det\big(\Sigma_i(\theta)\big)\}^{1/2}}$$

where $d_i = \#\{j : |j - i| \leq k\}$ is the dimensionality of $T_i(\mathbf{X})$, ranging from $1 + k$ to $2k + 1$ depending on if $i$ is close to the endpoints $i = 1$ and $i = m$, and $\mathbf{v}_i$ are $d_i$-dimensional row vectors. This gives

$$(3.11) \qquad \begin{aligned} &\pi_{m,i}(T_{m,i}(\mathbf{X}), \xi) \\ &= \frac{\sum_{T_i(\theta) \in \Omega_i^{(0)}} f_i\big(T_i(\mathbf{X}) \big| T_i(\theta)\big) w_0^{d_i - s_i(\theta)} (1 - w_0)^{s_i(\theta)}}{\sum_{T_i(\theta) \in \Omega_i} f_i\big(T_i(\mathbf{X}) \big| T_i(\theta)\big) w_0^{d_i - s_i(\theta)} (1 - w_0)^{s_i(\theta)}}, \end{aligned}$$

where $s_i(\theta) = \sum_{|j-i| \leq k} \theta_j$, $\Omega_i = \{0, 1\}^{d_i}$, and $\Omega_i^{(0)} = \{T_i(\theta) \in \Omega_i : \theta_i = 0\}$.

The estimation of the proportion of null hypotheses is an important aspect of the FDR problem. Benjamini and Hochberg [1] simply used the conservative $\widehat{w}_0 = 1$ to control the FDR at level $\alpha$, while Benjamini and Hochberg [2] suggested the possibility of power enhancement with estimated $w_0$ in the BH [1] procedure. Different estimators of $w_0$ were proposed by Storey [13] and Storey, Taylor and Siegmund [14] based on the tail proportion of $p$-values, and by Efron et al. [6] based on $\min_x\{f(x)/f_0(x)\}$, in the context of controlling the FDR. Tang and Zhang proved the consistency of (3.8) for normal mixtures.

## 4. Simulation results

In this section, we describe the results of our simulation study. We compare five procedures: the BH [1] rule using the true $w_0$, the approximate Bayes rule (2.6), and three EB rules (3.10) using $T_{m,i}(\mathbf{X}) = (X_j, |j - i| \leq 2)$, estimators (3.5), (3.6) and (3.7), and the following three values for (estimated) $w_0$: the true $w_0$, the Fourier estimator (3.8), and the bootstrap estimator (3.9). We denote the BH procedure by BH-$w_0$, and the three EB rules by EB-$w_0$, EB-Fourier and EB-bootstrap respectively. The target (conditional) FDR level is $\alpha = 0.1$ throughout the simulation study.

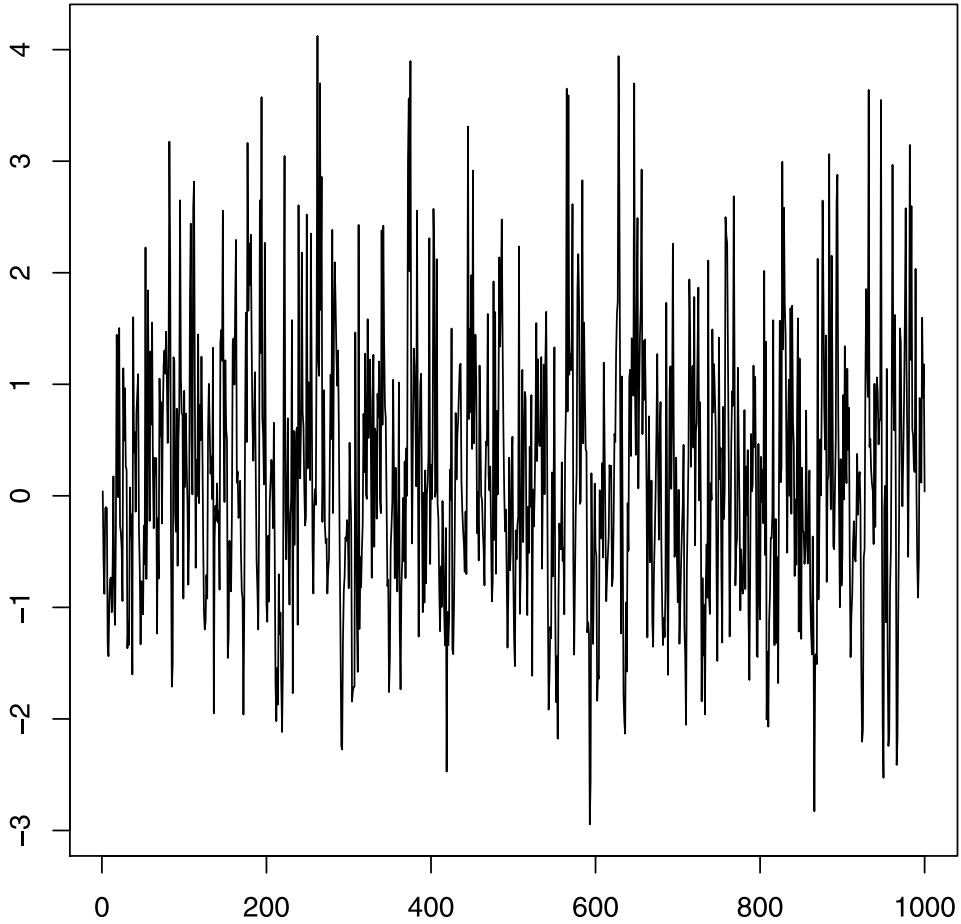

Fig 1. *A realization of the time series data.*

Simulation data are generated as follows: $m = 1000$,

$$\#\{i : \mu_i = 0\} = 900, \ \#\{i : \mu_i = 2\} = 100,$$

i.e. $w_0 = 0.9$, and $\gamma = (1, 0.6, 0.4, 0.2, 0.1, 0, 0, \ldots)$, e.g. $\gamma(1) = 0.6$. We note that this is a singular point in the nominal mixture model (3.3) for the derivation of EB procedures. A realization of the simulated $\mathbf{X}$ is plotted in Figure 1.

We plot the simulated pairs of $(V/R, R)$, i.e. the proportion of false rejections in the $x$-axis and the total number of rejections in the $y$-axis, for the BH-$w_0$ and the three EB procedures in Figure 2, with dashed lines at the means of the simulated data. The mean and standard deviation of $V/R$, $R$, and $V$ for all five procedures are given in Table 1. It is clear that the EB procedures are much more powerful than BH as they have much higher number $R$ of rejections, while the false discovery

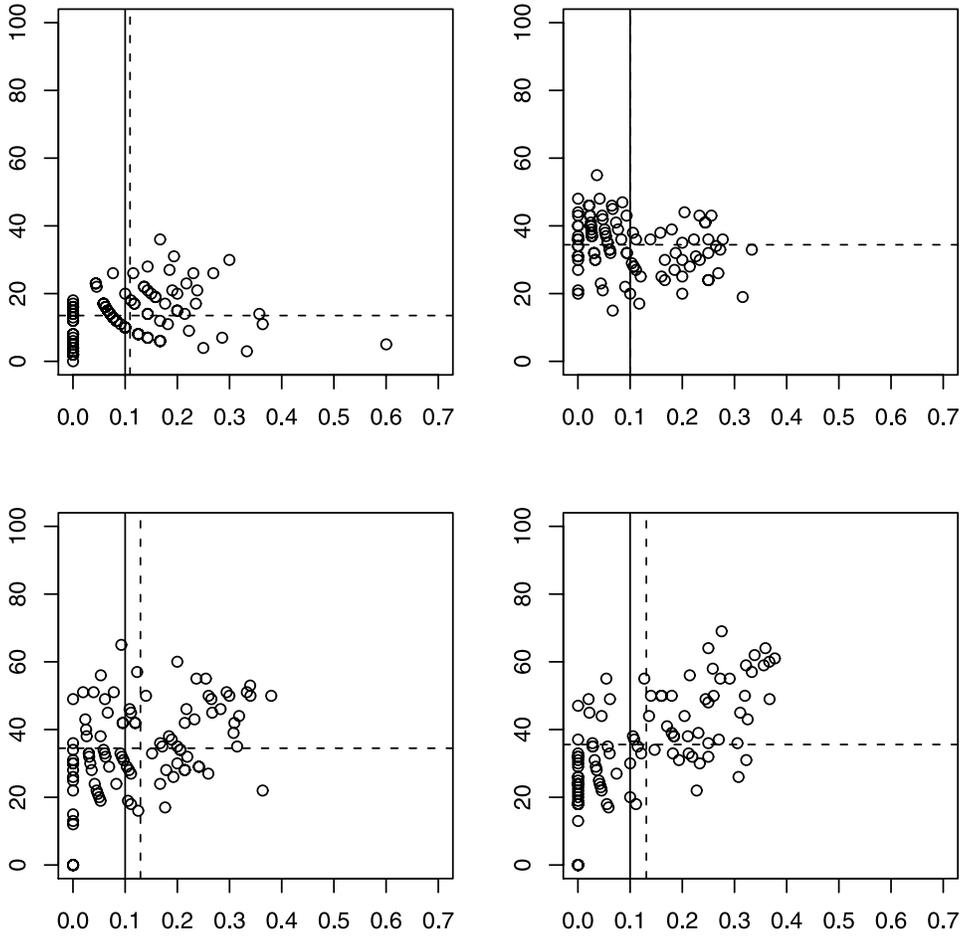

FIG 2. *The proportion of false rejections (x-axis) and the total number of rejections (y-axis) for BH-$w_0$, EB-$w_0$, EB-Fourier, and EB-bootstrap, clockwise from top-left; solid vertical lines for the target FDR of $\alpha = 10\%$ and dashed lines for the means of simulated points in the plots.*

ratio $V/R$ are similar among EB and BH procedures.

## 5. Conclusion

For multiple testing problems, we describe the Bayes and empirical Bayes approaches for controlling the (conditional) FDR proposed in Tang and Zhang [15]. While these approaches are completely general for dependent data, its implementation is subject to computational feasibility and the availability of sufficient information for the estimation of certain conditional probabilities of the individual null hypotheses. We propose in this paper to use the conditional probabilities of the null hypotheses given certain low-dimensional statistics to ease the computational burden and possibly to reduce the number of nuisance parameters involved. We implement this EB approach in a time series model with general stationary Gaussian errors. Simulation results demonstrate that the EB procedures have much high number of correct rejections and similar false rejection ratio compared with an application of the procedure of Benjamini and Hochberg [1] based on individual $p$- values. This clearly demonstrates the feasibility and superior power of the proposed EB approach for controlling the false rejection with dependent data.

Table 1
*Mean (μ) and Standard Deviation (σ) of V/R, R, and V*

| Procedure | V/R mean | V/R SD | R mean | R SD | V mean | V SD |
|---|---|---|---|---|---|---|
| BH-$w_0$ | 0.11 | 0.10 | 13.52 | 7.57 | 1.66 | 1.82 |
| Approximate Bayes | 0.12 | 0.03 | 76.86 | 6.62 | 9.12 | 3.16 |
| EB-$w_0$ | 0.10 | 0.09 | 34.42 | 7.95 | 3.27 | 3.05 |
| EB-Foureir | 0.13 | 0.11 | 34.49 | 13.44 | 5.06 | 4.95 |
| EB-bootstrap | 0.13 | 0.12 | 35.59 | 15.05 | 5.89 | 6.56 |